\documentclass[conference]{IEEEtran}
\usepackage[T1]{fontenc}
\usepackage{amsmath}
\usepackage[cmintegrals]{newtxmath}
\usepackage{url}
\usepackage{soul,color,graphicx,float,epstopdf}
\usepackage{tablefootnote,textcomp,gensymb}
\usepackage{eurosym,booktabs,multirow}
\usepackage[caption=false]{subfig}
\usepackage{amsfonts} 
\usepackage{algorithmic}
\usepackage{array}
\usepackage{stfloats}
\usepackage{float}
\usepackage{mathtools}
\usepackage{graphicx}
\usepackage{pdfpages}
\usepackage{subfig} 
\usepackage{balance}
\usepackage{xcolor}
\usepackage{comment}
\usepackage[normalem]{ulem}
\usepackage{dsfont}
\usepackage[linesnumbered, ruled]{algorithm2e}
\usepackage{enumitem}
\usepackage{tikz}
\usetikzlibrary{arrows,shapes,positioning,shadows,trees}

\usepackage{verbatim}
\usepackage{cite}
\usepackage{makecell}

\definecolor{Orange}{rgb}{1,0.5,0}

\newcommand*{\rom}[1]{\expandafter\@slowromancap\romannumeral #1@}

\usepackage{xcolor}
\usepackage{balance}

\begin{document}
%
\title{Integer-Based Pattern Synthesis for Asymmetric Multi-Reflection RIS}

\author{\IEEEauthorblockN{
Wei Wang\IEEEauthorrefmark{0},
Angela Doufexi\IEEEauthorrefmark{0},
Mark A Beach\IEEEauthorrefmark{0}
}                                     
\IEEEauthorblockA{\IEEEauthorrefmark{0}
Department of Electrical and Electronic Engineering, University of Bristol, Bristol, BS8 1QU, UK,\\Email: \{wei.wang, A.Doufexi, M.A.Beach\}@bristol.ac.uk}
}



\maketitle

\begin{abstract}
This study delves into the radiation pattern synthesis of reconfigurable intelligent surfaces (RIS) / reflection metasurfaces. Through superimposing multiple single-reflection profiles, which comprise the amplitude and/or phase settings of all constituent elements, a single incident wave can be effectively reflected in multiple asymmetric directions. However, some mismatch and interference between adjacent reflection beams may be caused by this superposition as well. Additionally, it is constrained by the inherent limitation that achieving linear and continuous amplitude adjustments and phase shifts in real-world designs is challenging. Consequently, the reconfigurable amplitude and phase must be approximated to discrete values, necessitating the arrangement of reflection profile before and after optimization based on integer. Therefore, in this paper, we adapt the traditional particle swarm optimization (PSO) algorithm to discretized integer-based PSO by proposing the concepts of 'discard rate' and 'knowledge.' With the enhancement of the integer-based programming, the multiple asymmetric reflection pattern can be synthesized with suppressed sidelobe levels within limited iterations and time cost.
\end{abstract}

\vskip0.5\baselineskip

\section{Introduction}
Reconfigurable intelligent surfaces (RIS) are emerging as a pivotal technology in the realization of future 6th-generation communications (6G) networks~\cite{di2020smart}. RISs offer low-cost cascaded channels that contribute to coverage enhancement~\cite{wu2019towards}, physical layer security~\cite{pan2021reconfigurable}, wireless power transfer~\cite{tran2022multifocus}, and numerous other applications. Among the components comprising the RIS system, the reconfigurable reflectarray (RRA) has evolved from the traditional reflectarray~\cite{berry1963reflectarray} through the incorporation of reconfigurable capabilities for each array element. This reconfigurable capabilities can be realized either actively~\cite{long2021active,liu2021active,tian2019design}, passively~\cite{li2020reconfigurable,yan2020passive}, or active-and-passive-jointly~\cite{song2020unsupervised}. Among them, passive reconfigurability comes out on top because of its low energy consumption and is thus more commonly adopted in RIS~\cite{elmossallamy2020reconfigurable}. Based on this passivity, the reflection performance of RIS, and also the multi-reflection superposition solely rely on the precise arrangement of each element's phase and amplitude, which we refer to as the ‘\textit{\textbf{reflection profile}}’ in this work. Therefore, the design of the reflection profile requires careful consideration and meticulous design.

Numerous methods and approaches have been proposed in recent years~\cite{nayeri2013design, zhao2021single, capozzoli2020cuda, aslan2019multiple, niccolai2020social} to improve the arrangement of reflection profiles, for either single-beam reflection~\cite{capozzoli2020cuda,niccolai2020social} or multi-beam reflection~\cite{nayeri2013design, zhao2021single, aslan2019multiple}, thereby enhancing performance or meeting special design requirements. Among them, \cite{nayeri2013design}~and~\cite{zhao2021single} modified the particle swarm optimization (PSO) to synthesize a single-feed multi-beam reflectarray and transmitarray, while \cite{capozzoli2020cuda} adopted a similar approach for single-feed single-beam reflection. Apart from PSO, convex optimization was also utilized for multiple beam synthesis in~\cite{aslan2019multiple}. In \cite{niccolai2020social}, a social network optimization (SNO)-based procedure for designing beam-scanning single-beam reflectarrays was proposed and compared with genetic algorithm~(GA) and PSO in the same model.
However, the majority of previous works have involved a time-consuming process, requiring as most as 44 hours~\cite{nayeri2013design}. This significant time consumption restricts the practical application of these synthesis methods in RIS, which necessitates real-time adjustment of the reflection profile. Meanwhile, the lengthy pre-training process is also unacceptable due to the time consumption of each reflect-direction.

As a matter of fact, the challenges associated with the arrangement of reflection profiles highly resemble integer programming. This is because the continuous adjustment of both amplitude and phase shift within RIS/RRA components is difficult to realize during practical manufacturing. Discretization and quantization are the essential processes to demonstrate RIS in reality.
The attainment of reconfigurability in RIS/RRA is reliant on the incorporation of electronically controlled devices into individual elements. However, the accurate translation between the control voltage and the desired amplitude/phase value is hard in real-world validation because of the hardware deviation. Besides, the system complexity and losses will also be increased observably with the resolution increment~\cite{theofanopoulos2020novel}.
Consequently, the majority of RRA hardware is limited to a resolution of 1-bit or 2-bit~\cite{yang20161,kamoda201160,yang20171600}, restricting their elements to operate within a mere 2 or 4 discrete amplitude and/or phase values. Nonetheless, by taking advantage of this hardware characteristic and structuring the optimization approach directly based on integer programming, we can still significantly enhance the efficiency of the optimization process.

\textit{Contributions:} In this work, we present an efficient and rapid pattern synthesis method for a single-incidence multiple-reflection RIS application with sidelobe suppression in undesired directions. The proposed method is a modification of PSO, where we innovatively introduce the concepts of '\textit{discard rate}' and '\textit{knowledge}' to transit PSO from continuous programming to discrete integer programming so that it can fit the optimization of reflection profile arrangement better. Moreover, the introduction of 'knowledge' significantly accelerates the optimization process, reducing the optimization time from several hours or even dozens of hours to a mere 8 minutes for any given single incidence angle and pair of reflection angles. With a completion time of minutes for each individual task, the overall elapsed time becomes acceptable when considering the pre-training required for incident and reflected directions across the entire space.

\textit{Paper outline:} Section~\ref{sec:Fundamental} expresses the antenna array synthesis equations and introduces the superposition method for multi-beam reflection related to our research. Then, section~\ref{sec:PSO} elaborates on our proposed integer programming PSO-based array synthesis in detail. This section includes the problem formulation in sub-section~\ref{subsec:formulation}, the concept of two proposed parameters in sub-section~\ref{subsec:Introducing the knowledge} and sub-section~\ref{subsec:Introducing the discard rate} respectively, and the final PSO setup in sub-section~\ref{subsec:Setup}. After that, the simulated results and analysis are shown in section~\ref{sec:Results}, and finally, section~\ref{sec:conclusion} concludes this paper.

\section{Multi-Reflection Synthesis Fundamental}
\label{sec:Fundamental}
\begin{figure}
\centering
\includegraphics[width=60mm]{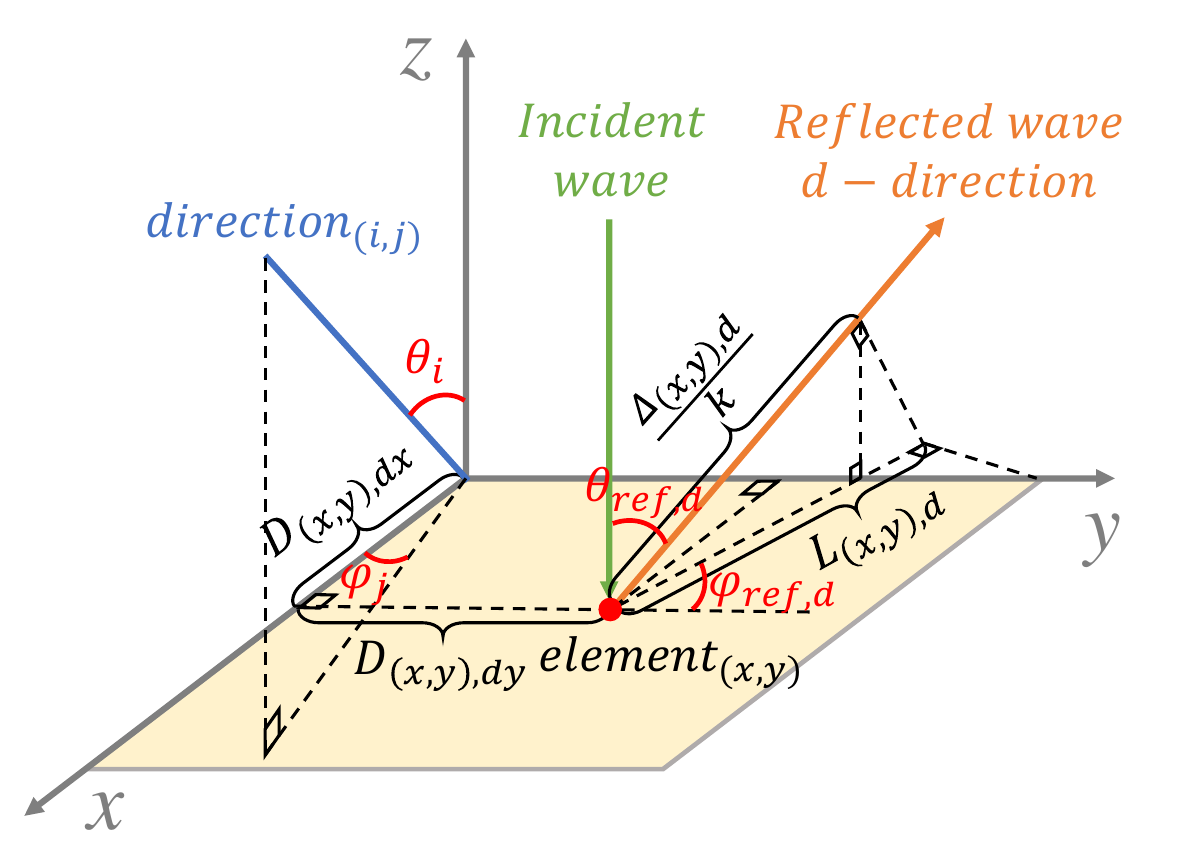}
\caption{The schematic diagram of the reflection pattern calculation.}
\label{fig:Synthesis}
\end{figure}
\IEEEpubidadjcol
For a more accurate emulation of real-life RIS application scenarios, where RRAs are deployed between base stations (BSs) and user equipment (UEs), the arrival angles of incident wave at each RRA element are set to be the same, corresponding to parallel incident waves. Additionally, to simplify the model, here we assume that the single-beam incidence is perpendicular to the given RRA surface as shown in Fig.~\ref{fig:Synthesis}. In this scenario, the phase compensation $\it{\Delta}_{(x,y),d}$ required for the RRA element at (\textit{x},\textit{y}) to achieve a reflected wave in \textit{d}-direction is:
\begin{equation}
    \it{\Delta}_{(x,y),d}=\it{k}\cdot L_{(x,y),d}\cdot sin\theta_{ref,d}\,,
\label{equi:1}
\end{equation}
where \textit{k} represents the wavenumber which is equal to $2\pi/\lambda$, and $\lambda$ is the wavelength of the incident wave in vacuum. $L_{(x,y),d}$ indicates the distance from $element_{(x,y)}$ to the array edge in \textit{d}-direction, while $\theta_{ref,d}$ is the included angle between the incident wave and the reflected wave in \textit{d}-direction.

After determining the actual phase difference, rounding should be made for the real-life RRAs due to the technical limitations in fabricating high-resolution RRA elements. The rounded value of $\it{\Delta}_{(x,y),d}$ is denoted as $\it{\Delta}^{*}_{(x,y),d}$ and depends on the designed resolution of the RIS. Based on the superposition principle for the generation of a multi-reflection profile, the phase compensation $\it{\Delta}_{(x,y), total}$ that the RRA $element_{(x,y)}$ should achieve can be expressed as:
\begin{equation}
    \it{\Delta}_{(x,y),total}=\frac{\sum_{d=1}^{D} \it{\Delta}^{*}_{(x,y),d}}{D}\,,
\label{equi:2}
\end{equation}
where \textit{D} is the total number of the reflected directions in a RIS multi-reflection scenario.
Then, the far-field intensity \textit{E} in $direction_{(i,j)}$ of the RRA can be calculated by:
\begin{equation}
    \it{E_{(i,j)}}={\bf E_{s}}\cdot\sum_{x={\rm 1}}^{X} \sum_{y={\rm 1}}^{Y} \beta_{(x,y)}\cdot e^{j[k\cdot OPD_{(x,y),(i,j)}+\Delta_{(x,y),total}]}\,,
\label{equi:farfield0}
\end{equation}
\begin{equation}
\begin{split}
    \it{OPD}_{(x,y),(i,j)}=&D_{(x,y),dx}\cdot {\rm cos}(\varphi_{j}){\rm sin}(\theta_{i})\\
    &+D_{(x,y),dy}\cdot {\rm sin}(\varphi_{j}){\rm sin}(\theta_{i})\,,
\end{split}
\end{equation}
where ${OPD}_{(x,y),(i,j)}$ represents the optical path difference of RRA $element_{(x,y)}$ in $direction_{(i,j)}$. $D_{(x,y),dx}$ and $D_{(x,y),dy}$ denote the distances of $element_{(x,y)}$ in x- and y-directions, respectively. $\bf{E_{s}}$ corresponds to the far-field pattern of a single element and is assumed to be isotropic here for simplified analysis. Additionally, $\beta_{(x,y)}$ is the amplitude of $element_{(x,y)}$ and is empirically set to be the constant 0.7 \cite{pei2021ris,araghi2022reconfigurable} because of the phased array we adopted.

However, multi-reflection synthesis generated from the superposition principle can only yield a coarse result. This limitation arises from the fact that although each single-reflection profile is optimal for its intended reflected wave to generate a reflected beam with low sidelobe levels (SLL), it may not be suitable for other reflected waves within the superposition. In some cases, these reflected waves may even require a completely inverse profile for their respective beams when two reflected directions are mirrored concerning the incident wave. Consequently, the superposition will eliminate the single-reflection profile of each other, resulting in the multi-reflection profile unfavorable for individual beams. Thus, further optimization is necessary to achieve a multi-reflection pattern with low SLLs.

\section{Integer Programming PSO-Based Synthesis}
\label{sec:PSO}
Particle swarm optimization \cite{kennedy1995particle} was proposed as a method to search for the global optimum by distributing numeral particles in a multi-dimensional hyper-volume. It is a commonly used optimizer in antenna array pattern synthesis \cite{khodier2005linear,bai2012hybrid,liao2020transmit} as well as reflectarray/transmitarray pattern synthesis \cite{nayeri2013design,zhao2021single,capozzoli2020cuda}. However, PSO was originally designed for nonlinear continuous programming. When applied to a multi-reflection RIS scenario where the reflection profile needs to be rounded to integers because of real-life resolution constraints, there can be a loss in optimization accuracy and an increase in computation time. Therefore, we propose two parameters: '\textit{\textbf{partical knowledge}}' and '\textit{\textbf{discard rate}}' to adapt the PSO algorithm suitable for nonlinear discrete (integer) programming and accelerate the optimization time.

\subsection{Pattern synthesis PSO formulation}
\label{subsec:formulation}
To initiate the PSO formulation for pattern synthesis of a sidelobe-suppressed multi-reflection RIS, we define the masks for the desired reflected wave in \textit{d}-direction ${\rm {\bf M}}_{W,d}\in \mathbb{C}^{I\times J}$ and the masks for all other directions ${\rm {\bf M}}_{U}\in \mathbb{C}^{I\times J}$ as:
\begin{equation}
{\rm {\bf M}}_{W,d}=\left\{
\begin{array}{cl}
1~,& (\theta_i-\theta_{ref,d})^2+(\varphi_j-\varphi_{ref,d})^2\leq R^2\,,\\
0~,& else\,.\\
\end{array} \right.
\label{equi:maskw}
\end{equation}
\begin{equation}
\quad {\rm {\bf M}}_{U}=\left\{
\begin{array}{cl}
0~,& (\theta_i-\theta_{ref,d})^2+(\varphi_j-\varphi_{ref,d})^2\leq R^2\,,\\
1~,& else\,.\\
\end{array} \right.
\label{equi:masku}
\end{equation}
where the reflected direction $d=1,2,...,D$, and $\varphi_{ref,d}$ is the included angle between the projection of \textit{d}-direction on the x-y plane and the y-axis. $R$ is set to be 10 degrees as experience value and $\mathbb{C}^{I\times J}$ is the matrix denoting the value range of $\theta_i$ and $\varphi_j$.
Then, map the discrete phase compensation values to positive integers starting from ‘1’. For example, if the RRA resolution is 2-bit, the phase compensations \{45°, 135°, 225°, 315°\} are mapped to \{1, 2, 3, 4\}.

In the PSO configuration, each particle is assigned a cognitive acceleration coefficient $c_1$ and social acceleration coefficient $c_2$, determining the balance between the influence of individual optimum and collective optimum on the particle's movement. In the meantime, an inertia factor $w$ is introduced to modulate the impact of a particle's past velocity on its current movement, affecting the exploring and exploiting ability of the particle in discovering both the global and local optima.
For the RRA with ${M\times N}$ elements, the velocity ${\rm {\bf v}}_p(t+1)\in \mathbb{C}^{M\times N}$ and position ${\rm {\bf x}}_p(t+1)\in \mathbb{C}^{M\times N}$ of particle~\textit{p} at the iteration~$(t+1)$ with respect to its current velocity ${\rm {\bf v}}_p(t)\in \mathbb{C}^{M\times N}$ and current position ${\rm {\bf x}}_p(t)\in \mathbb{C}^{M\times N}$ are updated by Eq.~(\ref{equi:velocity1}) and Eq.~(\ref{equi:position1}):
\begin{equation}
\begin{split}
    {\rm {\bf v}}_p(t+1):=~&w\cdot {\rm {\bf v}}_p(t)+c_1\cdot r_{1,p}\cdot ({\rm {\bf Pm}}_p(t)-{\rm {\bf x}}_p(t))\\
    &+c_2\cdot r_{2,p}\cdot ({\rm {\bf Gm}}(t)-{\rm {\bf x}}_p(t))\,,
\end{split}
\label{equi:velocity1}
\end{equation}
\begin{equation}
{\rm {\bf x}}_p(t+1):={\rm round}({\rm {\bf x}}_p(t)+{\rm {\bf v}}_p(t))\,,
\label{equi:position1}
\end{equation}
where $r_{1,p}$ and $r_{2,p}$ are two uniform random values generated within $(0,1)$ interval in order to increase the randomness of the search and they are different for each particle~\textit{p}. ${\rm {\bf Pm}}_p\in \mathbb{C}^{M\times N}$ is the cognitive best position of particle~\textit{p} while ${\rm {\bf Gm}}\in \mathbb{C}^{M\times N}$ is the social best position in history.

After achieving ${\rm {\bf v}}_p$, we impose a velocity restriction of $(-1,1)$ to the phase compensation of the elements, ensuring that their moving distances in each step remain within the range of $(0,1)$. Subsequently, as illustrated in Eq.~(\ref{equi:position1}), ${\rm {\bf v}}_p$ after the restriction is added with ${\rm {\bf x}}_p$, then rounded to the nearest integer due to the integer feature of resolution-limited RRA.
Suppose the resolution of RRA is \textit{K}-bit, so the number of phase compensation levels can be denoted as $2^K$. Based on it, the initial values of ${\rm {\bf v}}_p$ and ${\rm {\bf x}}_p$ are set as follows:
\begin{equation}
    {\rm {\bf v}}_p(t=0):=2\cdot{\rm {\bf rm}}_1-1\,,
\label{equi:velocity0}
\end{equation}
\begin{equation}
{\rm {\bf x}}_p(t=0):={\rm round}(2^K\cdot{\rm {\bf rm}}_2+0.5)\,,
\label{equi:position0}
\end{equation}
where ${\rm {\bf rm}}_1\in \mathbb{C}^{M\times N}$ and ${\rm {\bf rm}}_2\in \mathbb{C}^{M\times N}$ are random matrix with uniformly distributed numbers in $(0, 1)$. The ${\rm {\bf x}}_p(t=0)$ bigger than $2^K$ will be remapped to $2^K$ so that ${\rm {\bf x}}_p(t=0)$ will always be the integer within [1, $2^K$]. Meanwhile, the value range of ${\rm {\bf v}}_p(t=0)$ is set to be $(-1, 1)$.

By using the updated ${\rm {\bf x}}_p$ as the latest arrangement of phase compensation in Eq.~(\ref{equi:farfield0}), Eq.~(\ref{equi:farfield0}) can be rewritten as:
\begin{equation}
    \it{E_{p,(i,j)}}={\bf {E}_{s}}\cdot\sum_{x={\rm 1}}^{X} \sum_{y={\rm 1}}^{Y} \beta_{(x,y)}\cdot e^{j[k\cdot OPD_{(x,y),(i,j)}+{\rm {\bf x}}_{p,(x,y)}]}\,.
\label{equi:farfield1}
\end{equation}

Subsequently, combine Eq.~(\ref{equi:maskw}), Eq.~(\ref{equi:masku}), and Eq.~(\ref{equi:farfield1}), the performance of \textit{p}-particle at \textit{t}-iteration can be evaluated by:
\begin{equation}
\begin{split}
    SLL_p(t)=&~{\rm max}({\rm {\bf E}}_p(t)\cdot {\rm {\bf M}}_{U})\\
    &-{\rm min}\{\mathop{{\rm max}({\rm {\bf E}}_p(t)\cdot {\rm {\bf M}}_{W,d})}\limits_{\it{d={\rm 1},{\rm 2},\cdots,D}}\}\,.
\end{split}
\label{equi:SLL}
\end{equation}

During the optimization, the objection is to minimize the ${\rm max}\{SLL_p\}$~(${\it{p={\rm 1},{\rm 2},\cdots,P}}$) to determine the optimal phase arrangement of RRA. This optimization problem can be expressed as:
\begin{equation}
    {\bf P}:\quad{\rm minimize}\{{\rm max}\{\mathop{SLL_p}\limits_{\it{p={\rm 1},{\rm 2},\cdots,P}}\}\}.
\label{equi:Problem}
\end{equation}

\subsection{Introducing the knowledge}
\label{subsec:Introducing the knowledge}
The full possibility of phase arrangement for an RRA array with dimensions of ${M\times N}$ and the resolution of \textit{K}-bit is given by $(2^K)^{M\times N}$. However, due to the general size of RRAs~\cite{pei2021ris,araghi2022reconfigurable,liu2017concepts,yang20171600}, which typically consist of 100 to 3000 elements with 1-bit to 3-bit resolution, it is impractical to traverse all the possible arrangements. Additionally, using PSO-based synthesis starting from random positions would still take excessive time before sharp converging.

Therefore, we propose to introduce ‘knowledge’ which refers to the phase arrangement based on the empirical formulas, such as Eq.~(\ref{equi:2}) in section~\ref{sec:Fundamental}, into the initial ${\rm {\bf x}}_p(t=0)$. Here, we provide our definitions of ‘knowledge’ in this paper as below:
\begin{itemize}
\item \textbf{\textit{Zero-knowledge:}} random ${\rm {\bf x}}_p(t=0)$ for $p=1,2,...,P$~;
\item \textbf{\textit{Partial-knowledge:}} knowledge-based ${\rm {\bf x}}_1(t=0)$ and random ${\rm {\bf x}}_p(t=0)$ for $p=2,...,P$~;
\item \textbf{\textit{Full-knowledge:}} knowledge-based ${\rm {\bf x}}_p(t=0)$ for $p=1,2,...,P$~.
\end{itemize}

With the introduction of 'knowledge,' the optimization time is significantly reduced, allowing for rapid and efficient convergence. Furthermore, a comprehensive comparison of different ‘knowledge’ will be analyzed in Section~\ref{sec:Results}.

\subsection{Introducing the discard rate}
\label{subsec:Introducing the discard rate}
In the traditional PSO algorithm, the inertia factor $w$ was introduced to balance the globality of convergence and the converging speed~\cite{shi1998modified}, ensuring that the particles will not be restricted in the local optimums.
The regulation through the collaboration of $w$, $c_1$, and $c_2$ proves to be flexible and effective in continuous PSO optimization. However, its efficacy diminishes when applied to discrete PSO optimization problems. This limitation arises from the fact that the discrete optimizations only have a finite and discrete set of permissible value points. Therefore, the small $w$, $c_1$, and $c_2$ will make the particle adjustments in each step hidden in the round-off for the discrete values, meanwhile, the big $w$, $c_1$, and $c_2$ will still let the particles lost in the local optimums quickly.

As a consequence, we propose the 'discard rate' as a solution to address the converging balance problem within discrete PSO optimization for RIS applications. Specifically, we incorporate a cognitive discard rate $d_1$ and social discard rate $d_2$ to the cognitive and social components in Eq.~(\ref{equi:velocity1}) respectively. For each $d$, there is a 0-1 random matrix ${\rm {\bf d}}$ being generated with the probability of '0' to be $d$ and the probability of '1' to be ($1-d$).

\subsection{Final PSO setup}
\label{subsec:Setup}
After introducing the 'knowledge' and 'discard rate,' Eq.~(\ref{equi:velocity1}) and Eq.~(\ref{equi:position1}) can be converted to:
\begin{equation}
\begin{split}
    {\rm {\bf v}}_p(t+1):=~&w\cdot {\rm {\bf v}}_p(t)+c_1\cdot r_{1,p}\cdot {\rm {\bf d}}_1\cdot ({\rm {\bf Pm}}_p(t)-{\rm {\bf x}}_p^{*}(t))\\
    &+c_2\cdot r_{2,p}\cdot {\rm {\bf d}}_2\cdot ({\rm {\bf Gm}}(t)-{\rm {\bf x}}_p^{*}(t))\,,
\end{split}
\label{equi:velocity2}
\end{equation}
\begin{equation}
{\rm {\bf x}}_p^{*}(t+1):={\rm round}({\rm {\bf x}}_p^{*}(t)+{\rm {\bf v}}_p(t))\,,
\label{equi:position2}
\end{equation}
where ${\rm {\bf x}}_p^{*}(t=0)$ has different formation depending on the zero-, partial-, and full-knowledge which has been shown in detail in Section~\ref{subsec:Introducing the knowledge}. Meanwhile, the fitness function is defined based on Eq.~\ref{equi:SLL} as:
\begin{equation}
    {\bf Fitness}:\sum_{p={\rm 1}}^{P}(SLL_p).
\label{equi:fitness}
\end{equation}

For a better globality of convergence, we set the acceleration coefficients $c$, the discard rates $d$, and the inertia factor $w$ to be dynamically adjusted. The parameter values for these variables during various iteration stages are shown in Table~\ref{table:Table}. At the beginning, the particles undergo a near-random search with high $d$ and equal $c$. Then, they search with a higher weight of cognitive component at stage 2 to not lose the globality. At stages 3 and 4, they explore the potential optimum area and finally converge to the optimal value gradually.
The parameter value selection is based on our experience to make the proposed PSO converge within 100 iterations without losing the globality. Other parameter sets may also be applied to it.


\begin{table}
\centering
\renewcommand{\arraystretch}{1.5}
\caption{Parameters setup of proposed PSO}
\footnotesize
\label{table:Table}
\begin{tabular}{ccccc} 
\toprule
Parameters & Stage~1 & Stage~2 & Stage~3 & Stage~4\\ \midrule
$d_1$ & 0.8 & 0.4 & 0.2 & 0\\
$d_2$ & 0.8 & 0.6 & 0.2 & 0\\
$c_1$ & 1 & 1.2 & 1 & 0.9\\
$c_2$ & 1 & 0.8 & 1 & 1.1\\
$w$ & 0.6 & 0.4 & 0.2 & 0\\
\bottomrule
\end{tabular}
\renewcommand{\arraystretch}{1.00}
\end{table}

\section{Numerical Results and Analysis}
\label{sec:Results}

\begin{figure}
\centering
\includegraphics[width=85mm]{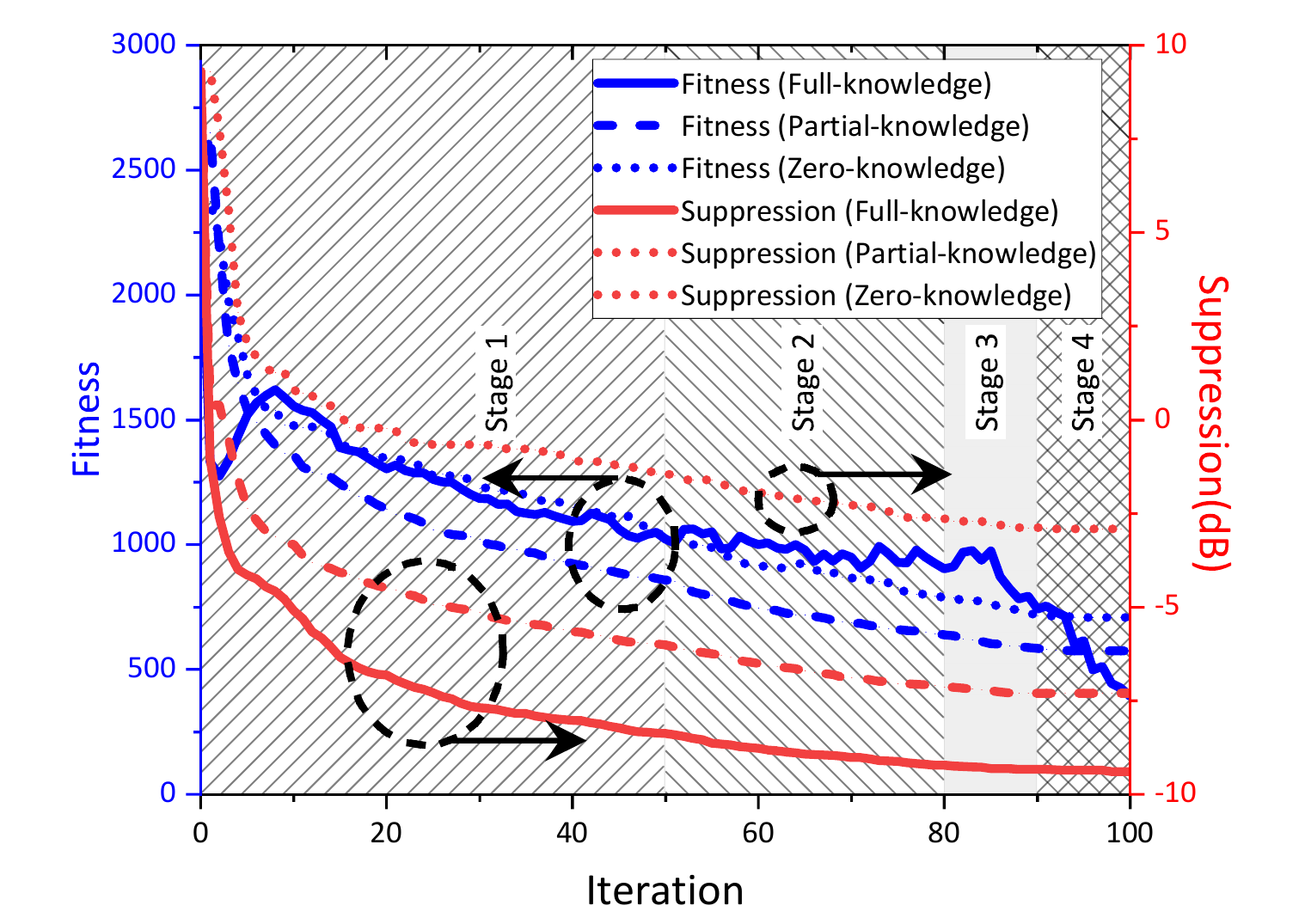}
\caption{Fitness and suppression performance of the proposed algorithm in a double-beam reflection scenario with 2-bit resolution. The performance comparison of zero-knowledge, partial-knowledge, and full-knowledge has been made.}
\label{fig:Knowledge}
\end{figure}
\vspace{-1.00mm}

We validate our proposed integer-based sidelobe suppression PSO algorithm through simulations conducted in a $30\times30$ RRA operating at 3.5~GHz. The inter-element spacing is set to be 21~mm which is around $1/4$ wavelength, and each element has a 2-bit resolution with four discrete and equal spacing phase values.
Fig.~\ref{fig:Knowledge} illustrates the performance of fitness and suppression in the scenario involving a single-perpendicular incident wave and two asymmetric reflection beams ($[\theta_{ref,1} = 45^\circ,~\varphi_{ref,1} = 30^\circ]$, $[\theta_{ref,2} = 45^\circ,~\varphi_{ref,2} = 110^\circ]$).
In this figure, the averaged curves from 20 independent simulations represent the outcomes derived from multiple validation trials. The blue lines represent fitness curves, while the red lines denote suppression curves.

Deriving from Eq.~(\ref{equi:Problem}), we employ the concept of '\textit{Suppression}' to express the '${\rm max}\{SLL_p\}$~(${\it{p={\rm 1},{\rm 2},\cdots,P}}$)'. This term quantifies the extent to which the sidelobe intensity in decibels (dB) falls below that of the expected main lobes. A positive value indicates that the sidelobes exhibit a higher intensity compared to the expected main beams.
When compared with the performance curves of zero-knowledge PSO in Fig.~\ref{fig:Knowledge}, both the partial-knowledge and full-knowledge exhibit significant improvements. Notably, in the case of partial-knowledge, only one particle possesses knowledge, whereas in the full-knowledge scenario, each particle possesses knowledge. Consequently, the full-knowledge curve consistently outperforms the partial-knowledge curve, particularly in the early iterations in Stage 1.
After our proposed integer-based PSO with full-knowledge, the '\textit{Suppression}' falls down from 0.4~dB to -9.6~dB, with suppression for about 10~dB, and the fitness curve also comes down synchronously.
The high fluctuations displayed in the fitness curve of full knowledge are from the prior knowledge of each particle. Some particles are initialized at the local optimums, making their SLLs undergo some degradations when moving from the local optimums to the global optimum.
Nevertheless, all these curves still undergo a sharp decline during stage 1 and a further decline during stages~2-4, revealing a good performance of our proposed algorithm and the effectiveness of the parameter selection in Table~\ref{table:Table}.

\begin{figure}
\centering
\includegraphics[width=87mm]{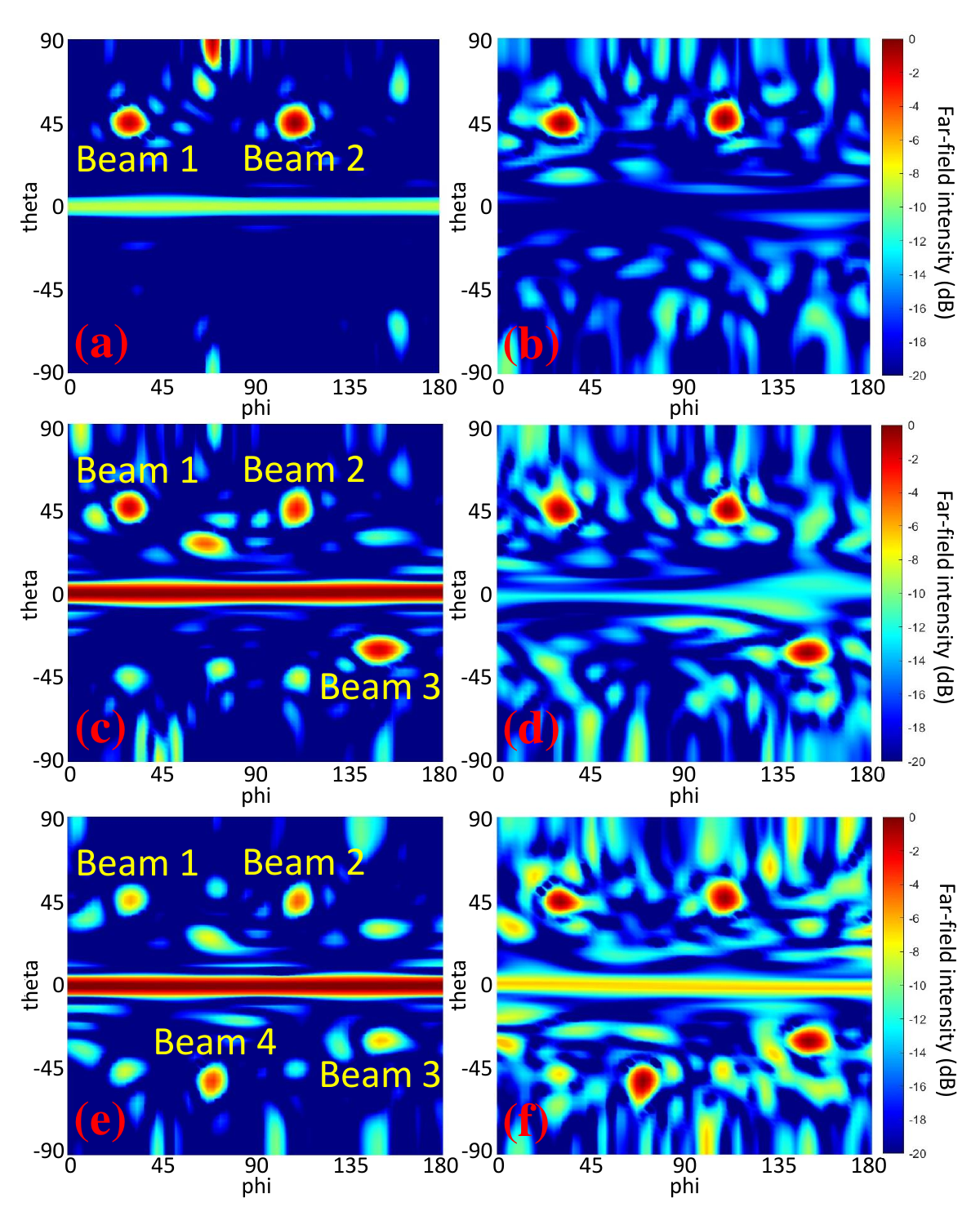}
\caption{The comparison of the multi-reflection RRA far-field patterns in directions: Beam-1 (45°, 30°); Beam-2 (45°, 110°); Beam-3 (-30°, 150°); and Beam-4 (-50°, 70°). (a) two-beam reflection before the optimization, (b) two-beam reflection after the optimization, (c) three-beam reflection before the optimization, (d) three-beam reflection after the optimization, (e) four-beam reflection before the optimization, and (f) four-beam reflection after the optimization.}
\label{fig:Multi-bit diagrams}
\end{figure}

Apart from the two-beam reflection, we also extend the validation of our proposed algorithm to multiple scenarios with different numbers of reflection beams. The simulation results of these validations are depicted in Fig.~\ref{fig:Multi-bit diagrams}, where we investigate the scenarios from two-beam reflection to four-beam reflection, pointing at asymmetric directions.
In Fig.~\ref{fig:Multi-bit diagrams}~(a), (c), and (e), the reflection patterns reveal the presence of high-intensity sidelobes in the asymmetric multi-reflections. These high-intensity sidelobes can lead to energy leakage and potential privacy disclosure. Notably, as the number of reflection beams increases, the impact of these sidelobes on the desired reflection beams becomes more pronounced. After the proposed optimization, the patterns in Fig.~\ref{fig:Multi-bit diagrams}~(b), (d), and (f) have a great improvement that the intensity of the lobes in unexpected directions have been suppressed for around 10~dB, making the intended reflection beams more obvious in the far-field pattern, much decrease the amount of energy and information leakage.

Notably, all the simulations mentioned above are based on our PSO algorithm with 100 particles. The execution time of it for a 100-iteration validation is approximately 8 minutes, significantly shorter than the previous works~\cite{nayeri2013design,capozzoli2020cuda,aslan2019multiple,niccolai2020social}. A detailed comparison can be found in Table~\ref{table:Comparison}, in which we define the '\textit{Optimization efficiency}' as:
\begin{equation}
\begin{split}
Optim.~efficiency~=\frac{Elements~\times~Individuals}{Optim.~time~\textit{[min.]}}\,.
\end{split}
\label{equi:efficiency}
\end{equation}

When compared to other works in this table, our work stands out for achieving an efficient multi-beam synthesis with sidelobe suppression within a limited time. Based on the Intel i7 desktop serial computing, the '\textit{optimization~efficiency}' of this work outperforms much higher than others, allowing the researchers to merge it into more complicated RIS applications, for example, the reconfigured multi-beam reflection.

\begin{table*}[t]
\centering
\renewcommand{\arraystretch}{2.2}
\caption{Comparison of different methods for RIS arrangement optimization}
\footnotesize
\label{table:Comparison}
\begin{tabular}{cccccccc} 
\toprule
Ref. (Year) & Beams & \makecell{Element\\number} & \makecell{Optimization\\algorithms} & \makecell{Individual\\number} & \makecell{Computing\\resources} & \makecell{Optimization\\time (minutes)} & \makecell{Optimization\\efficiency}\\ \midrule
TAP 2013 \cite{nayeri2013design} & Multiple & 848 & PSO & 400 & Serial & 2640 & 128\\
APS-URSI 2021 \cite{zhao2021single} & Multiple & 525 & PSO & 80 & Parallel & 9.6 & 4484\\
AEM 2020 \cite{capozzoli2020cuda} & Single & 2032 & PSO & 100 & Parallel & 120 & 1685\\
TAP 2020 \cite{aslan2019multiple} & Multiple & 64 & Convex & / & / & 360 & /\\
OJAP 2020 \cite{niccolai2020social} & Single & 144(576) & \makecell{SNO \\ PSO \\ GA} & 100 & Parallel & \makecell{621 \\ 613 \\ 613} & \makecell{23 \\ 23 \\ 23}\\
This work & Multiple & 900 & PSO & 100 & Serial & 8 & 11250\\
\bottomrule
\end{tabular}
\renewcommand{\arraystretch}{1.00}
\end{table*}

\section{Conclusion}
\label{sec:conclusion}
In this paper, we present an adaptation of the classical PSO into an integer-based PSO algorithm. By incorporating the concepts of 'knowledge' and 'discard rate' into each particle, our novel algorithm exhibits remarkable efficiency in addressing the multi-focus reflectarray pattern synthesis problem. We conducted a thorough performance comparison among 'zero-knowledge', 'partial-knowledge', and 'full-knowledge' in which the 'full-knowledge' based method outperforms the other two methods. Subsequently, we validate our proposed ‘full-knowledge’ integer-based PSO algorithm in scenarios involving reflected waves in asymmetric directions, including two, three, and four directions. Our simulation results demonstrate a substantial improvement of 10 dB in the suppression of SLLs for all the aforementioned scenarios.

\section*{Acknowledgments}
This work was partially supported by China Scholarship Council for funding the author under No.202108060224.
\bibliographystyle{IEEEtran} %
\bibliography{IEEEabrv,references} 
\end{document}